%% file: l4dc2024-sample.tex
\documentclass[12pt,final]{l4dc2024}
\DontPrintSemicolon
\usepackage{wrapfig}

\title[Deep Hankel matrices]{Deep Hankel matrices with random elements}
\usepackage{times}

 \coltauthor{\Name{Nathan P. Lawrence\nametag{\normalfont{ $\phantom{}^\pi$}}} \Email{input@nplawrence.com}\\
  \Name{Philip D. Loewen\nametag{\normalfont{ $\phantom{}^\pi$}}} \Email{loew@math.ubc.ca}\\
  \Name{Shuyuan Wang\nametag{\normalfont{ $\phantom{}^\mu$}}} \Email{antergravity@gmail.com}\\
  \Name{Michael G. Forbes\nametag{\normalfont{ $\phantom{}^\epsilon$}}} \Email{Michael.Forbes@Honeywell.com}\\
  \Name{R. Bhushan Gopaluni\nametag{\normalfont{ $\phantom{}^\mu$}}} \Email{bhushan.gopaluni@ubc.ca}\\
  \addr \nametag{\normalfont{$\phantom{}^\pi$}} Department of Mathematics, University of British Columbia \\
    \addr \nametag{\normalfont{$\phantom{}^\mu$}} Department of Chemical \& Biological Engineering, University of British Columbia \\
  \addr \normalfont{$\phantom{}^\epsilon$} Honeywell Process Solutions
  }

\usepackage{amsmath,amssymb,bm,bbm,amsfonts,mathtools} % math

\usepackage{shortex}
\usepackage{graphicx}
\graphicspath{{figures/}}

\usepackage{thmtools, thm-restate}

\usepackage{acro}
\acsetup{single=true}
\input{acronyms}

\begin{document}

\maketitle

\begin{abstract}%
Willems' fundamental lemma enables a trajectory-based characterization of linear systems through data-based Hankel matrices.
However, in the presence of measurement noise, we ask: Is this noisy Hankel-based model expressive enough to re-identify itself?
In other words, we study the output prediction accuracy from recursively applying the same persistently exciting input sequence to the model.
We find an asymptotic connection to this self-consistency question in terms of the amount of data.
More importantly, we also connect this question to the depth (number of rows) of the Hankel model, showing the simple act of reconfiguring a finite dataset significantly improves accuracy.
We apply these insights to find a parsimonious depth for \acs{LQR} problems over the trajectory space.
\end{abstract}

\section{Introduction}

This paper concerns Hankel matrices of the form
\[
H_{L}(z) =
\begin{bmatrix}
	z_{0} & z_{1} & \ldots & z_{N-L} \\
	z_{1} & z_{2} & \ldots & z_{N-L+1} \\
	\vdots & \vdots & \ddots & \vdots \\
	z_{L-1} & z_{L} & \ldots & z_{N-1}
\end{bmatrix},
\label{eq:Hankel}
\]
where each $z_i \sim \mathcal{N}(0,1)$.
This structure arises naturally in the context of a nominal \ac{LTI} system
whose state $x$ evolves in $\reals^n$:
\begin{equation}
\begin{aligned}
x_{t+1} &= A x_{t} + B u_{t} && \\
	\hat{y}_{t} &= C x_{t} && t=0,1,2,\ldots
\label{eq:LTI}
\end{aligned}
\end{equation}
By organizing the inputs and outputs of the system in Hankel matrices $H_L(u)$ and $H_L(\hat{y})$, respectively, Willems' fundamental lemma characterizes the trajectory space of \cref{eq:LTI} through the matrix-vector product $\begin{bsmallmatrix} H_L(u) \\ H_L(\hat{y}) \end{bsmallmatrix} \alpha$.
(A precise formulation is given in \cref{sec:background}.)
Taking the input to be a Gaussian probing signal $u$ and the output to be subject to measurement noise, $y = \hat{y} + \omega$, we arrive at the following situation:
\begin{equation}
	\begin{bmatrix}
		H_L(u) \\ H_L(y)	
	\end{bmatrix}
	\alpha
= 
\underbrace{
	\begin{bmatrix}
		H_L(u) \\ H_L(\hat{y})
	\end{bmatrix}
	\alpha}_{\text{True trajectory}}
	+ 
\underbrace{	
	\begin{bmatrix}
		0 \\
		H_L(\omega)
	\end{bmatrix}
	\alpha}_{\text{Error}}
\label{eq:noisymodel}
\end{equation}
The question then arises: which term dominates the right-hand side?
To give the problem a little more structure, we assume the input and output noise profiles are fixed, and we are only able to manipulate the depth and width, which are characterized by the window length $L$ and the number of samples $N$, respectively.
Therefore, we are interested in the interplay between the amount of data, the depth of the Hankel matrices, and the overall error.

\begin{figure}[tbp]
\centering
\includegraphics[width=8.4cm]{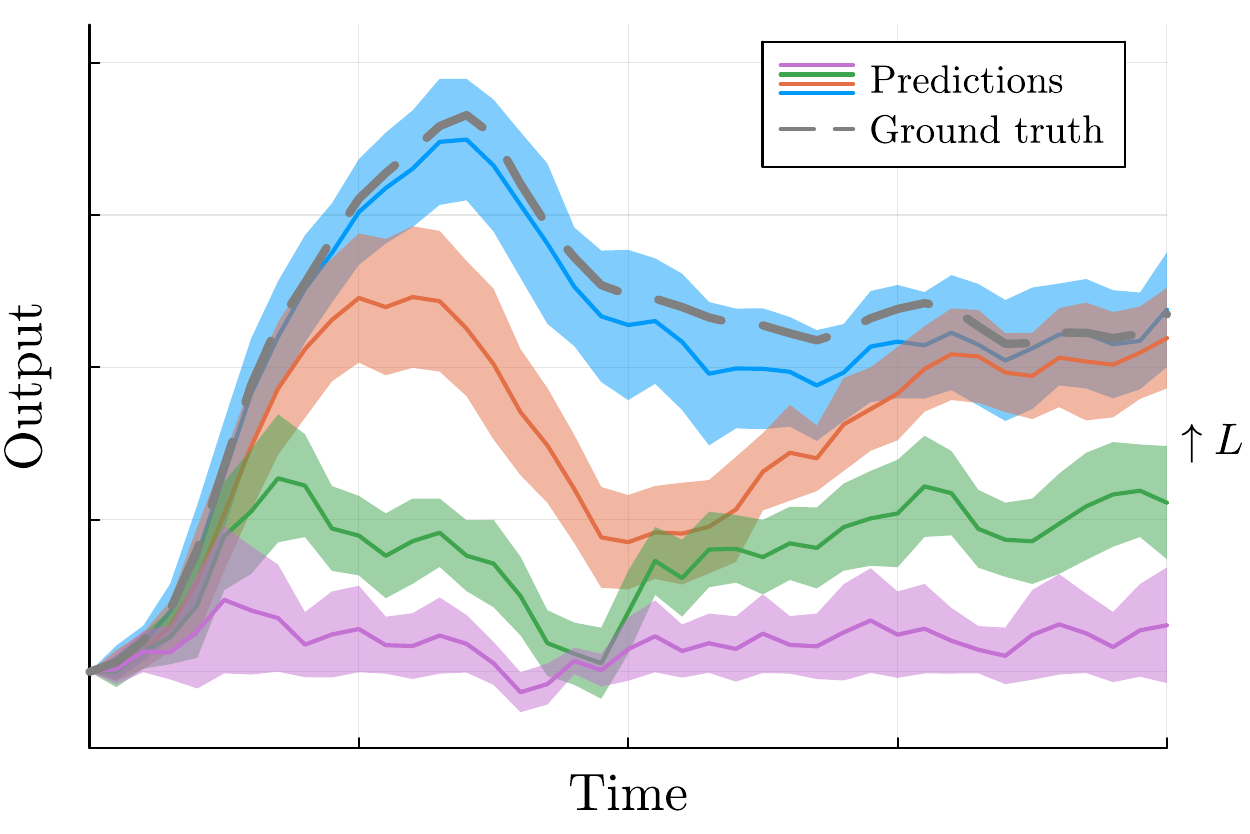}
\caption{For a fixed-size dataset, adjusting the depth of the input-output Hankel matrices dramatically improves self-consistency. Results are for $L=2,5,10,20$ and each color corresponds to $50$ rollouts with different output noise instances but a fixed input sequence.}
\label{fig:depthL}
\end{figure}

In practice, we employ a self-consistency test to determine a sufficiently expressive $L$ from a fixed dataset.
That is, we use the same input sequence $u$ in \cref{eq:noisymodel} to iteratively predict some output sequence $\tilde{y}$ with the hope that it is close to the underlying sequence $\hat{y}$.
We show in \cref{thm:randomL} and \cref{subsec:twoGaussians} that increasing the depth mitigates the effect of the noise term in \cref{eq:noisymodel}.
%We find that the parameter $L$ significantly influences the prediction accuracy. In other words, increasing the depth mitigates the effect of the noise term in \cref{eq:noisymodel}.
This is illustrated in \cref{fig:depthL} as motivation and explained in more detail in \cref{sec:results}.
%More precisely, given a noisy model of the form \cref{eq:noisymodel},

The most similar works to ours are \citet{coulson2023QuantitativeNotion,guo2023DataDrivenRobust,yan2023ApproximationSystem}.
However, none of them consider the matrix depth in their formulation.
\Citet{coulson2023QuantitativeNotion} propose casting Willems' lemma in terms of a minimum singular value criterion, rather than the standard binary rank condition.
\Citet{guo2023DataDrivenRobust} consider perturbations to the data-based model, similar to \cref{eq:noisymodel}, but their analysis is based on \emph{Page} matrices and the assumption that the perturbation has a known upper bound.
(We use \emph{Hankel} matrices and show the random perturbation in \cref{eq:noisymodel} is unbounded.)
\Citet{yan2023ApproximationSystem} examine the approximation properties of noisy Hankel models, but their analysis is framed in terms of independent rollouts rather than trajectory length and depth.
More related work is discussed in \cref{sec:related}.
Finally, the authors' work \citet{lawrence2023StabilizingReinforcement} contains similar analytical tools used here, but the setting is completely different: this work zooms in on the approximation properties of random Hankel matrices, while \citet{lawrence2023StabilizingReinforcement} uses the behavioral setting for reinforcement learning problems.

\section{Background}
\label{sec:background}

{\noindent\textbf{Notation.}} \quad
We often write a vector of sequential variables as 
$\bar{z} = \left[z_0, \ldots, z_k\right]\transpose$ 
when the number of elements is clear from context. When specifying the time indices, we write $\bar{z}_{0:k}$.
The \emph{spectral radius} function $\rho$ ingests a square matrix and returns a nonnegative scalar:
$\rho(M) = \max\left\lbrace \left\vert\lambda\right\vert\,:\, \lambda\in\mathbb{C},\ Mv = \lambda v\ \text{for some}\ v\ne0\right\rbrace.$
We use $\norm{\cdot}$ for the \emph{Euclidean norm} and $\norm{\cdot}_{F}$ for the \emph{Frobenius norm}.
$A^+$ denotes the \emph{Moore-Penrose inverse}, or \emph{pseudoinverse}, of the matrix $A$.

\noindent\textbf{Willems' fundamental lemma.}\quad We assume single-input single-output dynamics; however, the following formulation holds for general \ac{LTI} systems and multidimensional noise.
Given an $N$-element sequence $\{ z_{t} \}_{t = 0}^{N-1} \subset \reals$
and an integer $L$, $1\le L\le N$, the \emph{Hankel matrix of depth $L$} 
is the $L \times (N - L + 1)$ array with the constant skew-diagonal structure $H_L(z)$ defined in \cref{eq:Hankel}.
\begin{definition}
The signal $\{ z_{t} \}_{t = 0}^{N-1} \subset \reals$ is \emph{persistently exciting of order $L$} if $\rank(H_L(z)) = L$.
\label{def:PE}
\end{definition}

\begin{definition}
An input-output sequence $\{ u_{t}, y_{t} \}_{t = 0}^{N-1}$ is a \emph{trajectory} of \iac{LTI} system $(A, B, C)$ as in \cref{eq:LTI} if there exists a state sequence $\{ x_{t} \}_{t = 0}^{N-1}$ such that \cref{eq:LTI} holds.
\label{def:trajectory}
\end{definition}
We assume the matrices $A,B,C$ in \cref{eq:LTI} are unknown, $(A,B)$ is controllable, and $(A,C)$ is observable.
This lays the foundation for the following data-driven characterization of \ac{LTI} systems.
\begin{theorem}[Willems' fundamental lemma \citep{vanwaarde2020WillemsFundamental,willems2005NotePersistency}]
Let $\{ u_{t}, y_{t} \}_{t = 0}^{N-1}$ be a trajectory of \iac{LTI} system $(A, B, C)$ where $u$ is persistently exciting of order $L+n$. Then $\{ \bar{u}_{t}, \bar{y}_{t} \}_{t = 0}^{L-1}$ is a trajectory of $(A, B, C)$ if and only if there exists $\alpha \in \reals^{N-L+1}$ such that
\[\begin{bmatrix}
	H_L(u) \\
	H_L(y) 
\end{bmatrix}
\alpha =
\begin{bmatrix}
\bar{u} \\
\bar{y}	
\end{bmatrix}.
\label{eq:fundamentalLemma}\]
\label{thm:fundamentalLemma}%
\end{theorem}
Equivalently, one may check if the stacked matrix in \cref{eq:fundamentalLemma} has rank $L+n$ \citep{coulson2023QuantitativeNotion}.
\Cref{thm:fundamentalLemma} says that 
a Hankel matrix constructed from sufficiently exciting input-output data 
% spans all possible input--output trajectories of the underlying system. In other words, sufficiently exciting data are sufficient 
contains enough information to serve as a dynamic model.

The standard form above provides a certificate for a given trajectory; from that trajectory, one may also wish to advance it forward in time.
To that end, we use \cref{thm:fundamentalLemma} to generate arbitrarily long rollouts from some input sequence $\{ \hat{u}_t \}_t$ as follows:
Given $N$ and vectors $z_0,\ldots,z_N$, let 
$z = \{ z_{t} \}_{t = 0}^{N-1}$ and 
$z' = \{ z_{t} \}_{t = 1}^{N}$.
Then define the time-shifted Hankel matrix as follows:
\[
H'_{L}(z) = H_{L}(z').
\]
% Solve \cref{eq:fundamentalLemma} for $\alpha$, then assuming $L \geq n$, 
If $\alpha$ satisfies \cref{eq:fundamentalLemma}, then, assuming $L \geq n$,
multiplying $H'(y)$ by $\alpha$ is equivalent to advancing the internal state forward, yielding the unique next output trajectory $${\bar{y}' = H'(y) \alpha}.$$
 \Cref{alg:sim} repeats this procedure for any number of time steps.

\begin{algorithm2e}[tbp]
\caption{Data-driven rollout}\label{alg:sim}
\SetKwInput{KwInput}{Input}
\KwInput{Data $\{u_k, y_k\}_{k=0}^{N}$ with persistently exciting input of order $L+1+n$; Initial trajectory $\{\bar{u}_k, \bar{y}_k\}_{k = 0}^{L-1}$; An input sequence $\{\hat{u}_t \}$ for simulation.}
\For{each $\hat{u} \in \{\hat{u}_t \}$}{
	Solve for $\alpha$: $\begin{bsmallmatrix}
		H_L(u) \\
		H_L(y) 
	\end{bsmallmatrix}
	\alpha =
	\begin{bsmallmatrix}
	\bar{u} \\
	\bar{y}	
	\end{bsmallmatrix}$\;
	Compute the next element $\bar{y}' = H'_{L}(y)\alpha$\;
	Queue the next control input $\bar{u}_{L} = \hat{u}$\;
	Update trajectory: $\{\bar{u}_k, \bar{y}_k\}_{k = 0}^{L-1} \gets \{\bar{u}_k, \bar{y}_k\}_{k = 1}^{L}$\;
}
\end{algorithm2e}

%\subsubsection{Tools for analysis}

\noindent\textbf{Tools for analysis.} \quad Our problem setup postulates that the input sequence and output noise are (sub-)Gaussian.
To understand the approximation properties of noisy models such as \cref{eq:noisymodel}, we analyze the singular values of the purely noisy arrays introduced in \cref{eq:Hankel}.
Doing so is related to studying the norm of the pseudoinverse.
By extension, we can apply the techniques and results to the full noisy model in \cref{eq:noisymodel}.
In pursuit of this, we leverage two ingredients: 
$1$.~A Gershgorin disk theorem for generalized eigenvalue problems of the form $A v = \lambda B v$ \citep{nakatsukasa2011GerschgorinTheorem}. 
This gives bounds on the singular values of Hankel matrices using row-wise information.
$2$.~ The Hanson-Wright inequality \citep{rudelson2013HansonWrightInequality} for analyzing the concentrations of key terms that appear in item $1$.

\section{Main results}

We first analyze the singular values of random matrices of the form in \cref{eq:Hankel}. 
This will then give insight into its approximation properties of the full noisy model as a function of $N$ and $L$.

\subsection{Singular values of random Hankel matrices}
\label{subsec:singular}

%The following theorem establishes that the singular values of $H_L$ tend to infinity with high probability as the amount of data increases. 
The following theorem establishes that the singular values of $H_L = H_L(z)$ diverge to infinity with high probability as the amount of data increases.
\begin{theorem}	
For any $L \in \nats$, in the context detailed above, there is a sequence $\epsilon_N\to 0^+$ such that
\[
\lim_{N \to \infty} \Pr \left( \frac{1}{\sigma_{\min}(H_L)} \leq \eps_N \right) = 1.
\label{eq:probsingular}
\]
\label{thm:randomL}
\end{theorem}
See the appendix for a full proof.
The main idea is to employ a Gershgorin-type argument to contain all $L$ singular values
%$\frac{1}{\sigma^2}$ 
of $\left(H_{L}H_{L}\transpose\right)^{-1}$ within shrinking disks around the origin.
However, to make the problem tractable, we rewrite it as a generalized eigenvalue problem:
\[
I v = \frac{1}{\sigma^2} H_{L}H_{L}\transpose v
\label{eq:geneig}
\]
Therefore, we can employ concentration inequalities to analyze the diagonal and off-diagonal terms of $H_{L}H_{L}\transpose$ in a similar fashion to the classical Gershgorin disk theorem.
%Intuitively, the diagonal terms $\inner{\bm{z}_i}{\bm{z}_{i}}$ in \cref{eq:GramHankel} should grow much faster than the off-diagonal terms $\inner{\bm{z}_i}{\bm{z}_{j}}$.

The proof of \cref{thm:randomL} provides many suitable sequences $\epsilon_N\to 0^+$.
In particular, we arrive at the general estimate:
\[
\frac{1}{\sigma^2}
\le \frac{1}{(N-L+1)(1-\beta)}\left(1 + \frac{\gamma}{1-\gamma}\right),
\label{eq:estlambdai_preview}
\]
where ${\beta, \gamma \in (0,1)}$.
We therefore want $\beta, \gamma$ to be close to $0$.
For example, with a ``large'' ($N \gg L$) but finite dataset, setting $\beta = \gamma = \frac{1}{L+1}$ leads to the approximation  
\[
\frac{1}{\sigma} \lessapprox \frac{1}{\sqrt{N}}\frac{L+1}{L}.
\label{eq:scaleL}
\]
\begin{wrapfigure}[8]{r}{0.35\textwidth}
\vspace{-0.77cm}
\begin{center}
\includegraphics[width=0.35\textwidth]{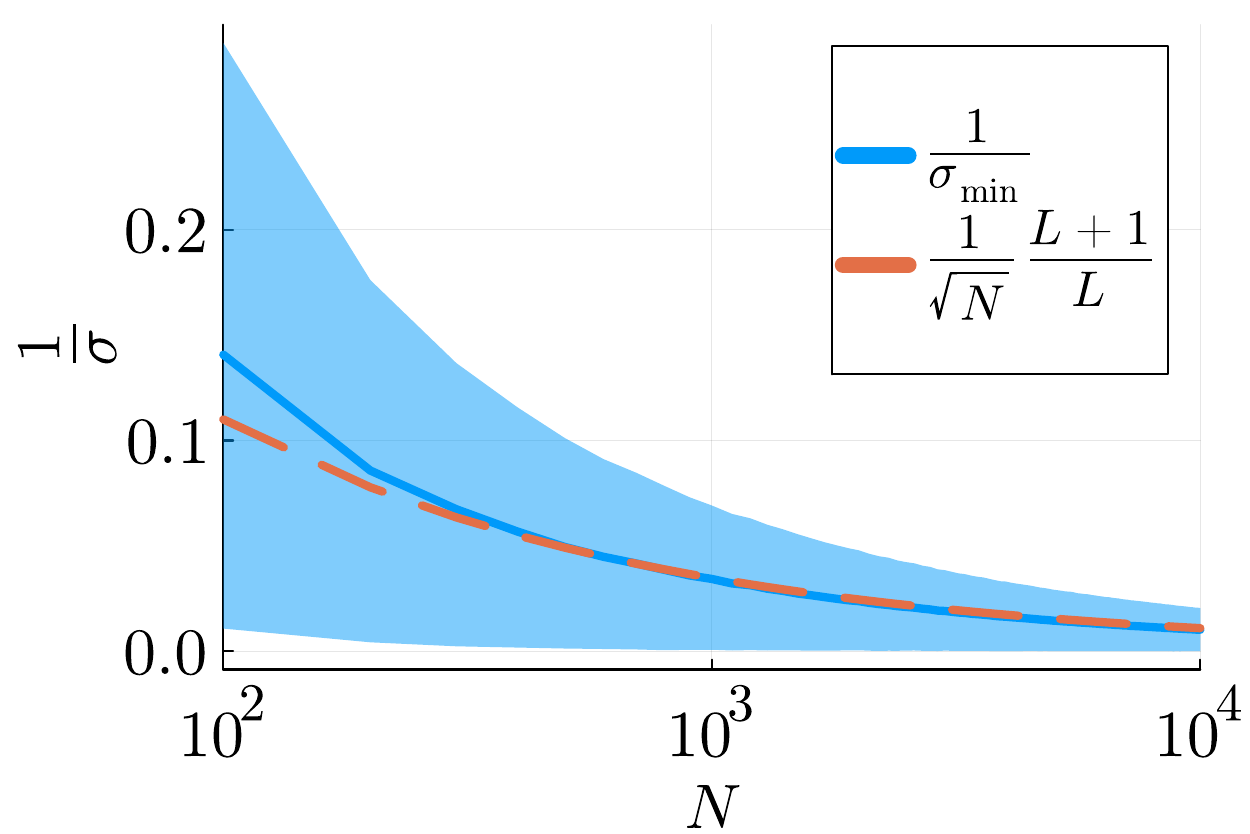}
\end{center}\vspace{-0.75cm}
\caption{Illustration of \cref{eq:scaleL}.}
\label{fig:singularnoise}
\end{wrapfigure}
Therefore, with a fixed amount of data, the bound on the singular values can be reduced by increasing $L$.
Of course, increasing $L$ is helpful only up to a point, as the quantity $\frac{L+1}{L}$ decreases to $1$ ``slowly'' meaning more data are preferable, if available.
Moreover, we caution that $L$ cannot be increased arbitrarily, as doing so decreases the underlying probabilities needed for \cref{eq:probsingular} to hold.
\Cref{eq:scaleL} is illustrated in \cref{fig:singularnoise} for a fixed $L$. The solid curve is the median inside the interquartile range across $50$ random instances of $H_L$.

\subsection{A tale of two Gaussians}
\label{subsec:twoGaussians}

We can apply the techniques described in \cref{subsec:singular} to characterize the singular values of the full noisy model in \cref{eq:noisymodel}.
In particular, we have two random matrices $H_L(u), H_L(\omega)$ and a third output component $H_L(\hat{y})$.
The input term $H_L(u)$ fits the formulation of the previous section and requires no modification.
We can leverage this fact to characterize the singular values of the full model.
Define $H_L$ to be the noisy stacked Hankel matrix in \cref{eq:noisymodel} and $\hat{H}_L$ to be the clean counterpart.

Putting these pieces together in a Gram matrix yields:
\[
\begin{multlined}
H_L H_L\transpose
=
\hat{H}_L \hat{H}_L\transpose \\
+ \begin{bsmallmatrix}
 0 & 0\\
 0 & H_L(\omega)H_L(\omega)\transpose 	
 \end{bsmallmatrix}
+
\begin{bsmallmatrix}
0 & H_L(u)H_L(\omega)\transpose	\\
H_L(\omega) H_L(u)\transpose & 0
\end{bsmallmatrix}
 + \begin{bsmallmatrix}
 0 & 0\\
 0 & H_L(\hat{y})H_L(\omega)\transpose 	+ H_L(\omega)H_L(\hat{y})\transpose 
 \end{bsmallmatrix}
\end{multlined}
\label{eq:GramData}
\]
To align with Willems' lemma, we are interested in the top $L+n$ eigenvalues of $H_L H_L\transpose$.
Using standard estimates, we can lower bound $\lambda_{L+n}(H_L H_L\transpose)$ by $\lambda_{L+n}(\hat{H}_L \hat{H}_L\transpose) + \sum_{i=1}^3 \lambda_{\min}(M_i)$
for each remaining matrix $M_i$ in \cref{eq:GramData}.
The first term, $\lambda_{L+n}(\hat{H}_L \hat{H}_L\transpose)$, comfortably tends to infinity using the Cauchy interlacing theorem \citep{coulson2023QuantitativeNotion,horn2012MatrixAnalysis} and \cref{thm:randomL}.
We can disregard the second term in \cref{eq:GramData}.
Finally, the last two terms threaten to shift the momentum away from $+\infty$.
The second-to-last term is concentrated around zero due to the classical Gershgorin disk theorem: one can obtain a well-behaved bound similar to \cref{eq:omij} in the proof of \cref{thm:randomL}.
The last term is similar if we assume a positive $c$ such that $\norm{\hat{y}_t} \leq c$ for all $t\geq 0$. (Otherwise, one expects to run into numerical stability issues when computing these singular values for large $N$ and a preconditioner should be used.)
Consequently, $\lambda_{L+n}\left( H_{L}H_{L}\transpose \right) \to \infty$ with high probability as in the case shown in \cref{subsec:singular}.

\subsection{Increasing depth improves self-consistency}

In light of the question surrounding \cref{eq:noisymodel}, \cref{thm:randomL} says the additive error term is not a ``small'' perturbation to the true data matrix.
However, this result still works in our favor.
Indeed, writing out the \emph{noisy} linear system implied by Willems' fundamental lemma (\cref{thm:fundamentalLemma}),
$H_L
\alpha =
\begin{bsmallmatrix}
\bar{u} \\
\bar{y}	
\end{bsmallmatrix}$,
the minimum-norm solution is 
$\alpha = H_L^+ \begin{bsmallmatrix}
	\bar{u} \\
	\bar{y}	
\end{bsmallmatrix}$. 
%Therefore, we are not interested in \emph{any} $\alpha$ in \cref{eq:noisymodel}; we are interested in the minimum-norm solution 
Consequently, our results and discussion from \cref{subsec:singular,subsec:twoGaussians} show that
\[
\lim_{N \to \infty} \Pr \Big( \underbrace{\norm{H_L^+}}_{\frac{1}{\sigma_{L+n}(H_L)}} \leq \underbrace{\eps_N}_{\to 0^+} \Big) = 1.
\]
Further, following \cref{eq:scaleL}, increasing $L$ accelerates the convergence by a linear factor.

To see the effect on the prediction accuracy, consider the following evaluation procedure:
Run \cref{alg:sim} where the simulation sequence $\{\hat{u}_t \} = \{u_k\}_{k=0}^{N}$.
Then, evaluate the \ac{RMSE} between the resulting outputs $\{\tilde{y}_t\}$ and the true measured sequence $\{y_k\}_{k=0}^{N}$.
Note each prediction has the general form $\tilde{y}_i = \left[y_L \ldots y_N\right] \alpha$ where $\left[y_L \ldots y_N\right]$ is the last row of $H'_L(y)$ and $y_{i + L}$ is the target.
Then, we see
\begin{align}
	\norm{\tilde{y}_i - y_{i + L}} 	&= \norm{\left(\left[\hat{y}_L \ldots \hat{y}_N\right] + \left[\omega_L \ldots \omega_N\right]\right) \alpha - y_{i + L}}\\
	&\leq \norm{\left[\hat{y}_L \ldots \hat{y}_N\right]\alpha - y_{i + L}} + \norm{\left[\omega_L \ldots \omega_N\right]\alpha}
\end{align}
We find that increasing $L$ has two desirable effects: 
$1$. It improves the bound on $\frac{1}{\sigma_{L+n}(H_L)}$ when $N  \gg L$, decreasing the minimum-norm solution.
$2.$ It decreases the number of noise terms on the right-hand side shown above.
Together, the influence of the noise term in the Hankel matrix $H_L(y)$ is mitigated.
In that spirit, balancing $N$ with a larger $L$ serves a similar function as a regularized solution by decreasing the norm of the $\alpha$ vector. A key difference, however, is that we are not introducing bias into the solution.

\section{More related work}
\label{sec:related}
%% I haven't seen a paper solely study the 	L parameter, but there's of course a lot of interest around
% solving the linear system of Willems lemma, so talk about that, the use of regularization, etc...conflicting consensus about L and N (less data/more data?), PCA can't work bc doesn't preserve Hankel structure, a priori knowledge of covariance, etc

The behavioral approach to control \citep{willems2005NotePersistency,markovsky2006exact,markovsky2008DatadrivenSimulation} has seen a resurgence in interest in recent years \citep{markovsky2021BehavioralSystems,martin2023GuaranteesDatadriven,faulwasser2023BehavioralTheory}.
This is largely due to the data-enabled predictive control (DeePC) framework introduced by \citet{coulson2019DataEnabledPredictive}.
\Citet{markovsky2021BehavioralSystems} gives an overview of the history, applications, and theoretical developments surrounding Willems' fundamental lemma, the key driver behind behavioral approaches to control.
In particular, significant attention has been given to ``robustifying'' Willems' result in the face of uncertainty \citep{depersis2020FormulasDataDriven,vanwaarde2023InformativityApproach} and its practical deployment \citep{huang2019DataEnabledPredictive,berberich2021DatadrivenModel}. See \citet{faulwasser2023BehavioralTheory} and the references therein for a recent account.

Measurement noise requires significant consideration, as Willems' lemma is squarely in the deterministic regime.
A standard approach is to regularize the solution $\alpha$ used for predictions \citep{coulson2019DataEnabledPredictive,markovsky2021BehavioralSystems,breschi2023DatadrivenPredictive}.
Other approaches take advantage of the structure of the uncertainty, leading to stochastic variants of the fundamental lemma \citep{pan2021StochasticFundamental, faulwasser2023BehavioralTheory} and maximum likelihood estimation techniques \citep{yin2020MaximumLikelihood,yin2023StochasticDataDriven} for dealing with measurement noise.
Further, robust control techniques, for example, based on the gap metric for uncertainty quantification or the $S$-lemma for robust stability, have been proposed \citep{padoan2022behavioral,vanwaarde2021NoisyData,berberich2023CombiningPrior}.

\section{Numerical experiments}
\label{sec:results}
%% 1. Experiment showing approximation accuracy over N and L
% compare against SSA, regularization, smoothing...
% finding smallest L such that L+k is only incrementally better
%% 2. LQR example over the \alpha dynamics for setpoint tracking

Code for the experiments is available: \url{https://github.com/NPLawrence/deepHankel}

\subsection{Rollout accuracy}
\label{subsec:ex_rollout}

\begin{figure}[tbp]
	\includegraphics[width=0.45\textwidth]{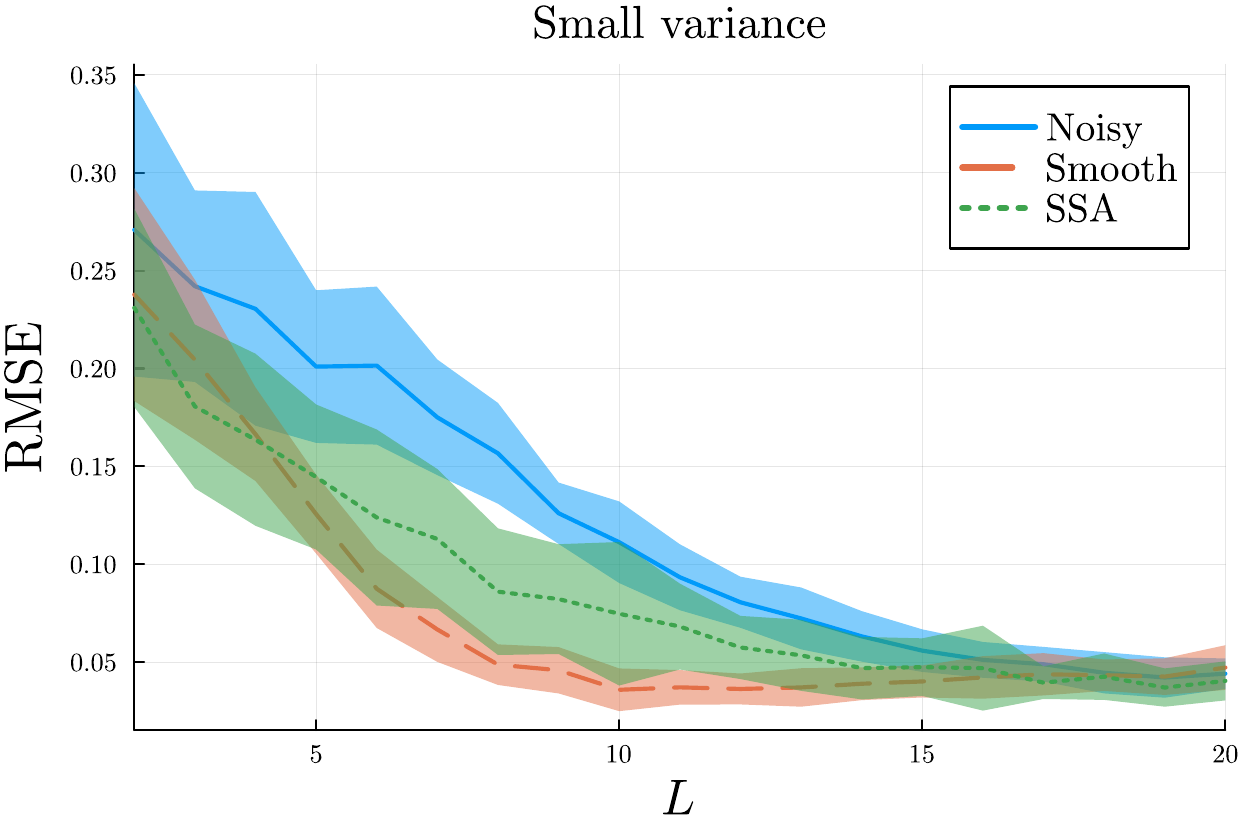}%
	\hfill%
	\includegraphics[width=0.45\textwidth]{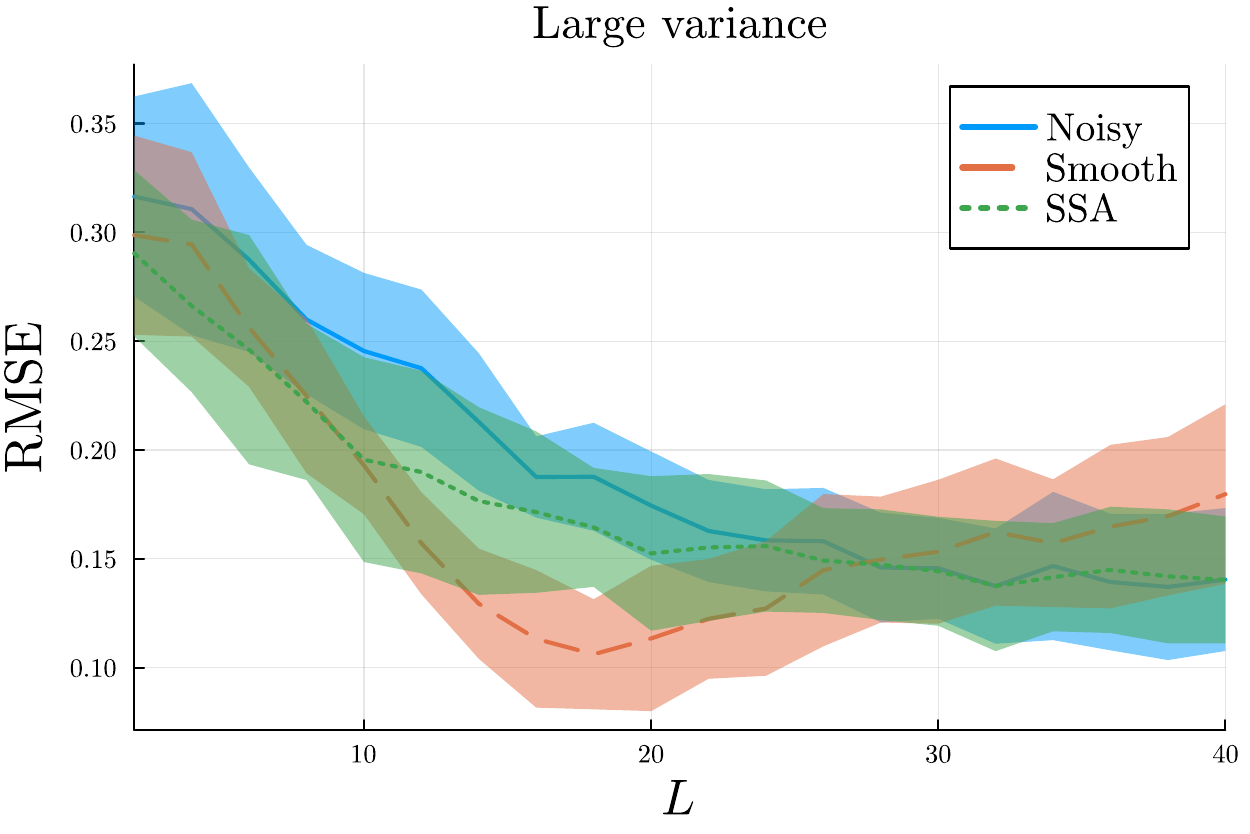}
	\caption{Data preprocessing versus simply using raw noisy data for evaluating self-consistency. The curves show the average \acs{RMSE} plus/minus the standard deviation over all rollouts.}
	\label{fig:RMSE}
\end{figure}

We revisit the motivating example in \cref{fig:depthL}:
Rollouts with small $L$ are nearly degenerate around the origin.
But with additional depth, $L$-length trajectories carry more of a ``trend'', hence becoming more distinguishable from noise and improving prediction accuracy.
For this example, we consider the system $\frac{1}{s^2 + 0.5 s + 1}$, discretized with a sampling time of $0.1$ seconds.
%The discretized system serves as the ``true'' underlying dynamics.
We assume a standard normal probing signal and output noise with variance $0.1$.

We repeatedly run \cref{alg:sim} as follows to gauge its prediction accuracy:
For a fixed $L$, the initial trajectory is set to the origin in $\reals^{2L}$.
Then for each ${N \in \{150, 200, 250\}}$, $10$ rollouts are performed under the same input sequence used in the Hankel matrix.
Each rollout uses a new sampled dataset to construct the Hankel matrices.
Finally, the \ac{RMSE} is calculated based on the rollout output trajectory and the true, noise-free underlying signal inside the dataset.
These \ac{RMSE} values are averaged, giving the results shown in \cref{fig:RMSE}.
We repeat this experiment with a noise variance of $1.0$ and ${N \in \{1500, 2500, 5000\}}$.

% In \cref{fig:RMSE}, multiple curves are on display.
The curves in \cref{fig:RMSE} illustrates the \ac{RMSE} as a function of $L$.
%This is to further illustrate the effect of the $L$ parameter.
We repeat the above procedure using $3$ preprocessing steps of varying complexity:
\begin{itemize}
\item {\textbf{Do nothing:}}\quad This is referred to as ``Noisy'' and corresponds to using the raw sampled data.
\item {\textbf{Smoothing:}}\quad This is referred to as ``Smooth'' and corresponds to replacing the measured input--output data with a moving average:
	\[
	\left\{u_t, y_t\right\}_{t=L-1}^{N-1} \leftarrow \left\{\frac{1}{L} \sum_{k = t-L+1}^{t} u_{k}, \frac{1}{L} \sum_{k = t-L+1}^{t} y_{k}\right\}_{t=L-1}^{N-1}
	\label{eq:smooth}
	\] 
	$L$ serves both as the depth parameter of the Hankel matrix and the moving average window length.
	With no noise, this is a realizable trajectory, as it corresponds to averaging the internal state sequence.
	When multiplying the resulting averaged Hankel matrix by some $\alpha$, this strategy is effectively an ensemble of time-shifted noisy Hankel matrices.
\item {\textbf{\Ac{SSA} \citep{hassani2007singular}:}}\quad 	\Ac{SSA} is a time series analysis method similar to PCA with the key feature of preserving the Hankel structure. Since a reconstructed matrix with the top $L$ singular values/vectors is not necessarily a Hankel matrix, \ac{SSA} ``Hankelizes'' it through skew-diagonal averaging, thereby recovering a new signal.
\end{itemize}

All three preprocessing methods lead to a steep descent in \ac{RMSE} with respect to $L$. 
Smoothing the data has the most dramatic descent, both with small and large output noise. However, the error starts to increase after depth $20$ in the right plot of \cref{fig:RMSE}:
While \cref{eq:smooth} decreases the variance of the output noise, it also decreases the magnitude of the input excitation; moreover, this strategy essentially removes $L$ additional columns from the model (relative to the other methods), tampering with the $L-N$ balance alluded to in \cref{subsec:singular}.
\Ac{SSA} is also a reasonable option in both cases. However, in the large variance setting, it is nearly indistinguishable from the ``Noisy'' strategy since a large $L$ corresponds to recreating the noise.

In all cases, the simple act of increasing the depth from a fixed dataset has a dramatic positive effect on rollout performance.
This is true even without any preprocessing, as illustrated in \cref{fig:depthL}.
However, there are only incremental gains after a certain point.
The next example shows that finding the beginning of this plateau is also useful for control.

\subsection{Application to LQR}

We formulate the \ac{LQR} problem over the space of trajectories and study the influence of depth on closed-loop performance.
Starting with the standard linear equation$\begin{bsmallmatrix}
H_L(u) \\ H_L(y)	
\end{bsmallmatrix}
\alpha =
\begin{bsmallmatrix}
\bar{u} \\
\bar{y}	
\end{bsmallmatrix},
\label{eq:willems}$
we note that by sequentially applying new inputs $\hat{u}$, \cref{alg:sim} relates the current solution vector $\alpha$ to the solution at the next time step $\alpha'$ as follows:
\begin{equation}
\left[
\begin{array}{c}
H_L(u) \\ 
H_L(y)	
\end{array}\right] \alpha'
= 
\left[
\begin{array}{c}
	H'_{L-1}(u) \\ \hline 
	0 \\ \hline
	H'_L(y)
\end{array}\right]
\alpha
+
\left[
\begin{array}{c}
0 \\ \hline
1 \\ \hline
0
\end{array}\right]
\hat{u}
=
\left[
\begin{array}{c}
\bar{u}_{0:L-2} \\ \hline
\hat{u} \\ \hline
\bar{y}'
\end{array}\right].
\label{eq:trajLQR}
\end{equation}
That is, \cref{eq:trajLQR} describes the evolution from $\bar{u}, \bar{y}$ to $\bar{u}', \bar{y}'$.

Standard \ac{LQR} solvers can be readily applied to the trajectory-space model in \cref{eq:trajLQR}.
In particular, assuming a pseudoinverse solution, one obtains a feedback controller of the form $u = -K \alpha,$ which can be rewritten as
\[
u = -K \left[
\begin{array}{c}
H_L(u) \\ 
H_L(y)	
\end{array}\right]^{+}
\begin{bmatrix}
\bar{u} \\
\bar{y}	
\end{bmatrix}.
\]
This feedback controller only uses previous input-output data, rather than employing explicit state estimation, to perform actions.
We deploy this idea for setpoint tracking on the plant
\[
P(z) = 0.1159\frac{z^3 + 0.5z}{z^4 - 2.2z^3 + 2.42z^2 - 1.87z + 0.7225}
\]
studied by \citet{ljung1999model, pillonetto2010NewKernelbased, yin2020MaximumLikelihood}.
Integral action is achieved by defining a state-space model around the variable $x = \begin{bsmallmatrix}
 	\alpha' - \alpha \\
 	r - y
 \end{bsmallmatrix},$ where $r$ is a reference signal, and solving for the optimal gain matrix. (See, for example, \citet{young1972ApproachLinear}.)

\begin{figure}[tbp]
	\includegraphics[width=0.31\textwidth]{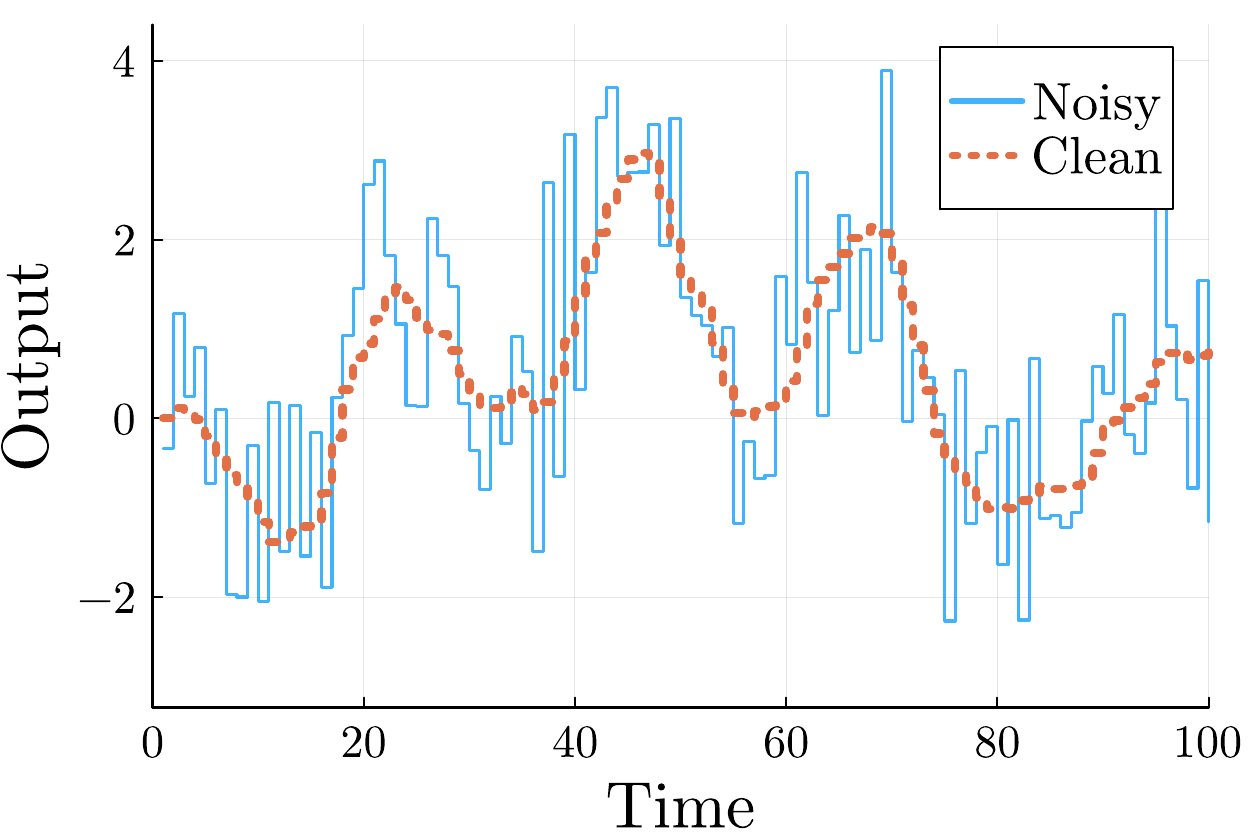}%
	\hfill%
	\includegraphics[width=0.31\textwidth]{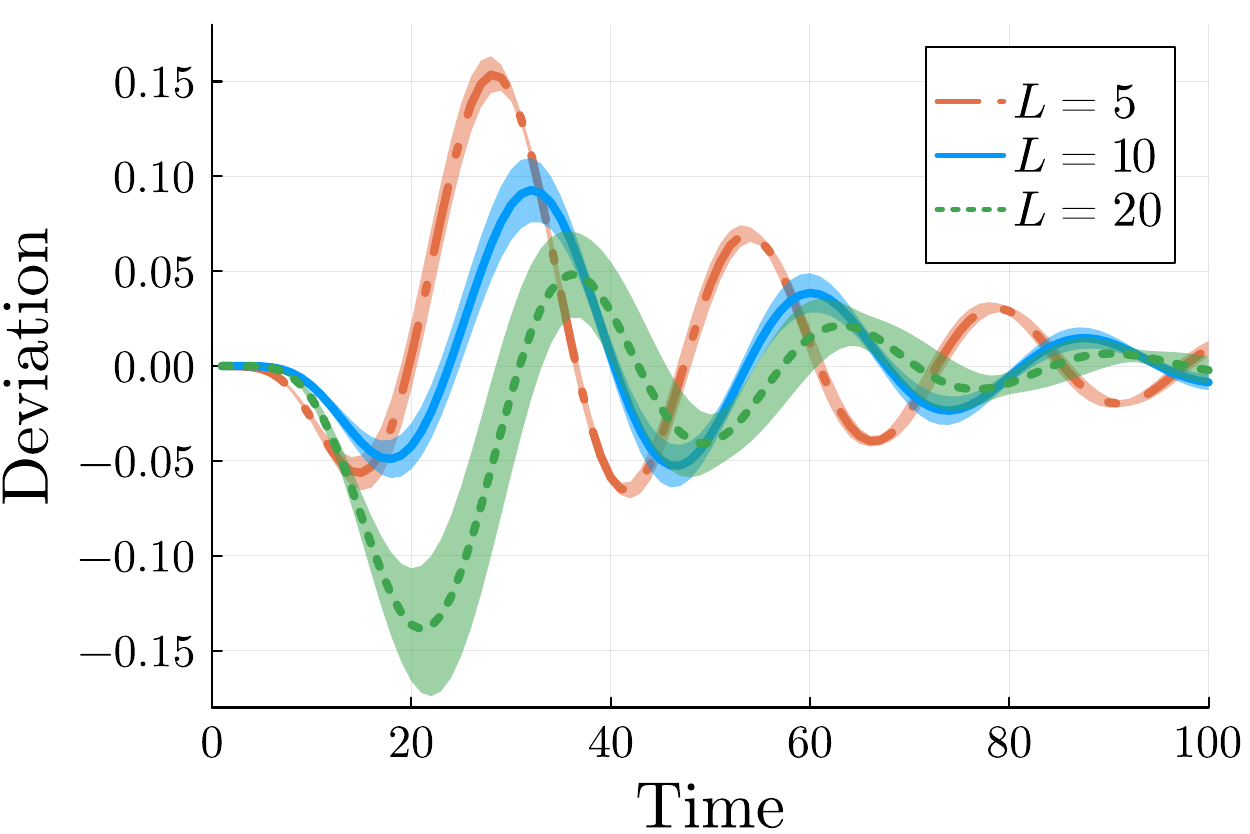}%
	\hfill%
	\includegraphics[width=0.31\textwidth]{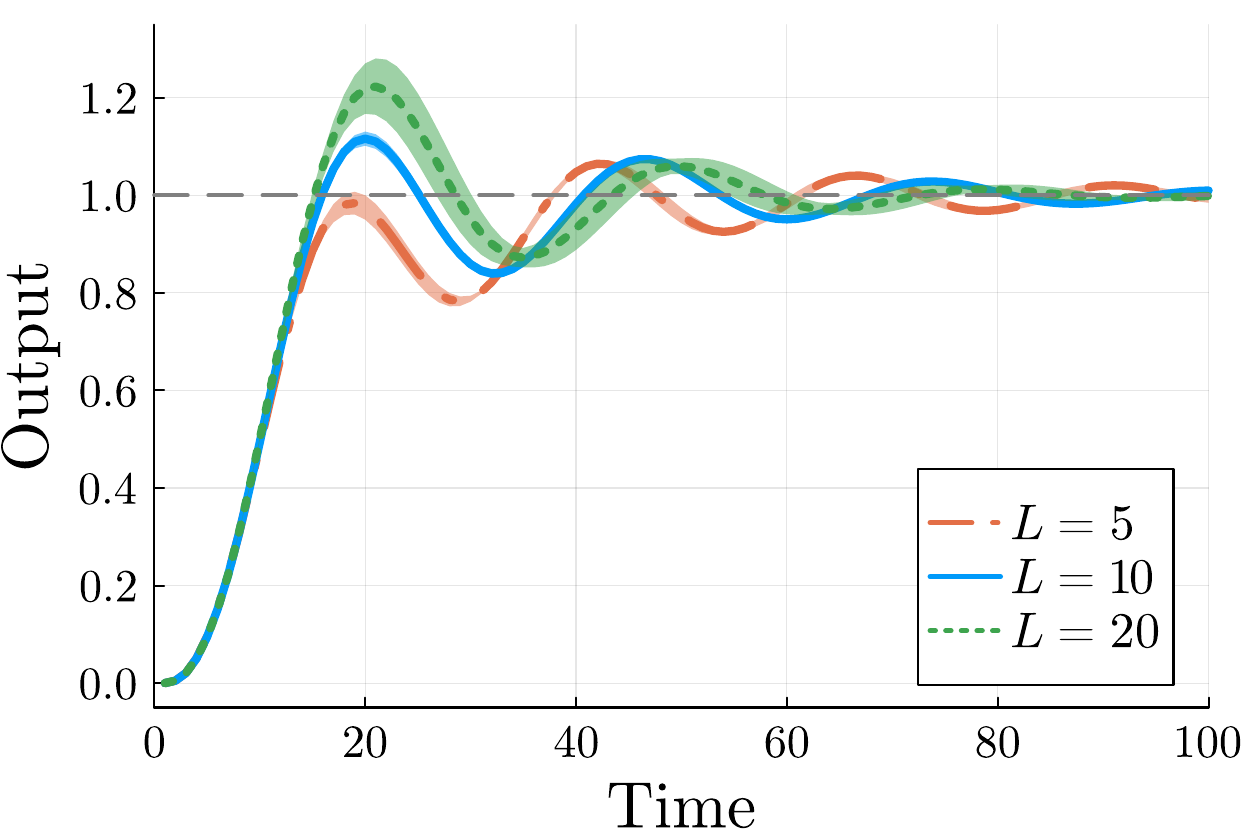}
	\caption{(left) Noisy output data; (center) The difference between the closed-loop trajectory from using a noisy model to obtain \iac{LQR} controller and the noise-free counterpart; (right) Tracking performance of the noisy controller on the true system.}
	\label{fig:lqr}
\end{figure}

We collect $400$ input-output samples from $P$ with standard normal input and output noise.
Some of these samples are shown in \cref{fig:lqr}.
This collection of data is used to run the self-consistency test from \cref{subsec:ex_rollout}.
$L = 10$ is a minimal yet expressive depth, as it marks the beginning of a plateau akin to that in \cref{fig:RMSE}.
To illustrate this, we perform $3$ LQR experiments with $L = 5,10,20$.
\Cref{fig:lqr} shows that $L = 10$ deviates the least from the ground-truth closed-loop trajectory.

\section{Conclusion}
%% Things about impulse response (why more data would be bad),
% extensions to DeePC, nonlinear systems with some static nonlinearity, or time-varying component

We have illustrated the effectiveness of deep Hankel matrices for approximation and control.
In particular, we showed that the length of an input-output trajectory is important for \emph{asymptotic} approximation properties, while depth influences the \emph{transient} behavior of this rate.
Practically, for a long rollout of excitation data, simply reconfiguring the Hankel matrices for a parsimonious depth has a profound impact on performance.
However, we caution that this would not hold in other settings, such as with an impulse response.
Therefore, future work should study different probing profiles and extend these results to systems with static nonlinearities or time-varying elements.
%Moreover, we have focused on stable single-input single-output systems for simplicity; however, \cref{thm:fundamentalLemma} is general, and \cref{thm:randomL} can be extended to multidimensional noise.

\section*{Proof of \cref{thm:randomL}}
\label{sec:appendix}

Inventing the special bold font notation for row vectors with $N-L+1$ consecutive components 
\[
\bm{z}_i = 
\begin{bmatrix}
	z_i & z_{i+1} & \ldots & z_{i+N-L}
\end{bmatrix}
,\qquad i=0,1,\ldots,L-1
\label{eq:defomegai}\]
gives the following convenient notation for the 
$L\times L$ Gram matrix that often appears in what follows:
$H_{L}H_{L}\transpose = \left[ \inner{\bm{z}_i}{\bm{z}_{j}} \right]_{i,j=1:L}$
where $H_L = H_L(z)$.
Let $z_0, z_1, \ldots,$ be a sequence of IID standard normal random variables.
Fix $N\ge L$; for convenience, let $\hat N = N-L+1$ denote the number of
components in the row vectors $\bm{z}_i$ defined in~\cref{eq:defomegai}.

We are interested in the set of nonnegative $\sigma$ for which
$H_{L}H_{L}\transpose v = \sigma^2 v$ for some real-valued vector $v \neq 0$, or equivalently written in \cref{eq:geneig}.
This generalized eigenvalue problem is suitable for a Gershgorin-type argument by containing all $\frac{1}{\sigma^2}$ in a set of disks around the origin. 
In particular, we utilize Corollary $2.6$ in \citet{nakatsukasa2011GerschgorinTheorem}: By  formulating conditions under which $H_{L}H_{L}\transpose$ is diagonally dominant, it follows that all $\frac{1}{\sigma^2}$ are captured by the union of $L$ disks:
\[
\frac{1}{\sigma^2} \in \bigcup_{i = 0}^{L-1} \left\{ s \in \comps \colon \abs{s - c_i} \leq \rho_i \right\},
\label{eq:disks}
\]
where the centers and radii are defined by
$
	c_i 
= \frac{1}{\inner{\bm{z}_{i}}{\bm{z}_{i}}}$ and $\rho_{i}= \frac{r_i}{\inner{\bm{z}_{i}}{\bm{z}_{i}}(1 - r_i)}  
$, respectively,
with $r_i = \frac{1}{\inner{\bm{z}_{i}}{\bm{z}_{i}}} \sum_{j \neq i} \abs{\inner{\bm{z}_{i}}{\bm{z}_{j}}}$ for $i=0,\ldots,L-1$.

To harness this fact, fix any constants $\beta,\gamma$ in $(0,1)$ and introduce
\begin{equation}
\theta = \frac{\gamma(1-\beta)\hat N}{L-1},
\quad\text{so}\quad
0 \le (L-1)\theta = \gamma(1-\beta)\hat N.
\label{eq:alphatheta}
\end{equation}
Consider the random event in which all of the following inequalities hold:
\begin{align}
&\abs{\inner{\bm{z}_i}{\bm{z}_i} - \hat N} \le \beta\hat N,
&&i=0,1,\ldots,L-1;
\label{eq:omii}\\
&\abs{\inner{\bm{z}_i}{\bm{z}_j}} \le \theta,
&&i,j=0,1,\ldots, L-1\ \text{with}\ i\ne j.
\label{eq:omij}
\end{align}
These conditions suffice to make $H_{L}H_{L}\transpose$ strictly diagonally dominant,
because (for each $i$)
\[
\inner{\bm{z}_i}{\bm{z}_i}
\ge \hat N - \beta\hat N
= \gamma^{-1} (L-1)\theta
> \sum_{j\ne i}\abs{\inner{\bm{z}_j}{\bm{z}_i}}.
\]
For each $i$, inequality~\cref{eq:omij} and definition~\cref{eq:alphatheta} imply
$r_i \le \frac{(L-1)\theta}{\hat N(1-\beta)}= \gamma.$
We have the bound on the center and radius
\begin{equation}
	\abs{c_i} \le \frac{1}{\hat N(1-\beta)}
\qquad\text{and}\qquad
\rho_i \leq \frac{\gamma}{\hat{N}(1-\beta)(1-\gamma)}.
\label{eq:estri}
\end{equation}
Applying \cref{eq:disks} for the disks with indices $i=0,\ldots,L-1$, this implies
\[
\frac{1}{\sigma^2}
\le \abs{c_i} + \rho_i
\le \frac{1}{\hat{N}(1-\beta)}\left(1 + \frac{\gamma}{1-\gamma}\right).
\label{eq:estlambdai}\]
Therefore, the right side of~\cref{eq:estlambdai} converges to $0$ as $N \to \infty$.
Define $\eps_N$ by equating $\eps_N$ with the right side of~\cref{eq:estlambdai}.
Then~\cref{eq:estlambdai} says precisely that $\rho\left(\left(H_{L}H_{L}\transpose\right)^{-1}\right)\le \eps_N$,
and we have just shown that $\eps_N\to 0$ as $N\to\infty$.
To complete the proof, it remains only to estimate the probabilities of
the random events in \cref{eq:omii,eq:omij} in terms of $N$.

This is a consequence of the Hanson-Wright inequality \citep{rudelson2013HansonWrightInequality}, which establishes the existence of a universal constant $c > 0$ such that
\[
\Pr\left( \abs{z\transpose  M z - \EE \left[ z\transpose  M z \right]} > t \right) \leq 2 \exp\left( -c \min \left\lbrace \frac{t^2}{\norm{M}_{F}^{2}}, \frac{t}{\norm{M}} \right\rbrace \right),\qquad t\ge0.
\label{eq:HW}
\]
for any matrix $M$
Indeed, consider the $\hat{N}$-dimensional row vectors $\bm{z}_i$, $i=0,\ldots,L$.
Each one can be extracted from the extra-long row $z=[z_0\ \ldots\ z_N]$ 
by multiplication with a suitable block-structured matrix:
$\bm{z}_i = z U_i$ where $U_i = \begin{bsmallmatrix}
 	0 \\ I \\ 0
 \end{bsmallmatrix}
$ has $i$ zeros in the top block, $L-i$ zeros in the bottom block, and $I \in \reals^{\hat{N}\times\hat{N}}$.
Thus we have
$\inner{\bm{z}_i}{\bm{z}_j}
= \bm{z}_i{\bm{z}_j}\transpose
= z U_i {U_j}\transpose z\transpose.$
Each $M_{ij}=U_i{U_j}\transpose$ is a zero matrix of size $(N+1)\times(N+1)$
containing an embedded $\hat N\times\hat N$ identity matrix.
The embedding puts the $\hat N$ nonzero elements of matrix $M_{ij}$ on the diagonal if and only if $i=j$.
Thus we have $\EE \left(z M_{ii} z\transpose\right) = \hat N$ for $0\le i\le L$,
and $\EE \left(z M_{ij} z\transpose\right) = 0$ whenever $i\ne j$.
Now \cref{eq:HW} gives
\begin{align}
\Pr\left( \abs{\inner{\bm{z}_{i}}{\bm{z}_{i}} - \hat{N}} \le \beta\hat N \right) 
&\ge 1 - 2 \exp\left(-c \beta^2 \hat{N}\right) &&\to 1 &&& 0\le i\le L-1,\\
\Pr\left( \abs{\inner{\bm{z}_{i}}{\bm{z}_{j}}} \le \theta \right) 
&\ge 1 - 2 \exp\left( -c\frac{\gamma^2 (1-\beta)^2 \hat{N}}{(L-1)^2} \right) &&\to 1 &&&i \neq j,\ 0\le i,j\le L-1.
\label{eq:HWterms}
\end{align}

In summary, by choosing $N$ sufficiently large, \cref{eq:omii,eq:omij} define $L^2$ random events that hold with probability arbitrarily close to $1$.\hfill $\square$

% Acknowledgments---Will not appear in anonymized version
\acks{We gratefully acknowledge the financial support of the Natural Sciences and Engineering Research Council of Canada (NSERC) and Honeywell Connected Plant.
We would also like to thank Professor Yaniv Plan for helpful discussions.}

\bibliography{2024_l4dc}

\end{document}

%% file: acronyms.tex
\DeclareAcronym{RL}{short = RL, long = reinforcement learning, short-indefinite = an}
\DeclareAcronym{IQC}{short = IQC, long = integral quadratic constraint}
\DeclareAcronym{MPC}{short = MPC, long = model predictive control}
\DeclareAcronym{LQR}{short = LQR, long = linear quadratic regulator, short-indefinite = an}
\DeclareAcronym{LTI}{short = LTI, long = linear time-invariant, short-indefinite = an}
\DeclareAcronym{BIBO}{short = BIBO, long = {bounded-input, bounded-output}}
\DeclareAcronym{SISO}{short = SISO, long = {single-input, single-output}}
\DeclareAcronym{MIMO}{short = MIMO, long = {multiple-input, multiple-output}}
\DeclareAcronym{SVD}{short = SVD, long = singular value decomposition}
\DeclareAcronym{PID}{short = PID, long = proportional-integral-derivative}
\DeclareAcronym{PI}{short = PI, long = proportional-integral}
\DeclareAcronym{YK}{short = YK, long = \YK}
\DeclareAcronym{RMSE}{short = RMSE, long = root-mean-square error}
\DeclareAcronym{SSA}{short = SSA, long = singular spectrum analysis}